\documentstyle[epsfig,graphics,graphicx,preprint,aps]{revtex}
\begin{document}
\draft

\title{The de Broglie-Bohm Interpretation 
of Evaporating Black-Holes}

\author{J. Acacio de Barros
\footnote{Email: acacio@fisica.ufjf.br
}, G. Oliveira-Neto
\footnote{Email: gilneto@fisica.ufjf.br
} and T. B. Vale
\footnote{Email:tiberio@fisica.ufjf.br
}}

\address{Departamento de F\'{\i}sica, 
Instituto de Ciencias Exatas, Universidade
Federal de Juiz de Fora, CEP 36036-330, 
Juiz de Fora, Minas Gerais, Brazil.}

\date{\today{}}
\maketitle

\begin{abstract}
In this work we apply the de Broglie-Bohm interpretation of quantum
mechanics to the quantized spherically symmetric black-hole coupled
to a massless scalar field. The wave-functional used was first obtained
by Tomimatsu using the standard ADM quantization and a gauge that
places the observer close to the black-hole horizon. Using the causal
interpretation, we compute quantum trajectories determined by the
initial conditions. We show that the quantum trajectories for the
black-hole mass can either increase or decrease with time. The quantum
trajectories that show increasing mass represent the usual black-hole
behavior of continuous energy absorption. The mass-decreasing quantum
trajectories are a purely quantum mechanical phenomena. They can be
physically interpreted as describing a black-hole that evaporates. 
\end{abstract}
\pacs{03.65.Ta, 04.60.Ds, 04.60.-m, 97.60.Lf}

\section{Introduction.}

\label{sec:introduction}

Since the fundamental discovery, made by S. W. Hawking, that 
black-holes may emit radiation \cite{hawking}, many studies 
have been made in order to better understand this process. 
Initially, most of the works
were concentrated in the area of quantum field theory in 
curved space-time
\cite{wald}. More recently, some physicists have started 
studying
the Hawking radiation with the aid of a quantum gravity 
theory \cite{tomimatsu},
\cite{kiefer}, \cite{casadio}, \cite{vaz}. Most of these works
deal with the theory of quantum general relativity. In this theory,
the standard probabilistic Copenhagen interpretation of quantum 
mechanics
cannot be applied. New interpretations of quantum mechanics have been
proposed over the years to deal with quantum general relativity 
\cite{everett},
\cite{griffiths}. Another interpretation of quantum mechanics that
may be applied to quantum general relativity is the causal 
interpretation.

The causal interpretation of quantum mechanics was first proposed
by de Broglie, and later on it was extended by Bohm to include 
many-particle
systems and fields \cite{bohm}. In this interpretation, variables
corresponding to observable physical quantities have an onthological
meaning regardless of whether they are observed or not, contrary to
the standard Copenhagen interpretation of quantum mechanics.

The problems of applying Copenhagen's interpretation of quantum 
mechanics
to quantum cosmology has raised recent interest on the causal 
interpretation
of quantum mechanics in quantum cosmology \cite{acacio}, as this
interpretation does not need an external observer to bring a 
observable
into reality. The causal interpretation has been applied with 
success,
by several authors, to quantum general relativity \cite{acacio},
\cite{vink}.

In the present work we would like to study the Hawking radiation 
process
using the theory of quantum general relativity and the causal 
interpretation.
We shall use the wave-functionals derived in Tomimatsu's work 
\cite{tomimatsu}.
There, he considered a spherically symmetric space-time, minimally
coupled to a massless scalar field. He wrote the Hamiltonian form
of the theory and derived from it the supermomentum and the 
superhamiltonian
constraints. In order to derive these constraints he used a 
particular
gauge that places the observer close to the black-hole horizon. 
Tomimatsu
found two wave-functionals by solving the operatorial version of the
superhamiltonian constraint. The first wave-functional was interpreted
as representing the classical black-hole behavior, mainly because
the expected value of the time derivative of the black-hole mass is
positive. The second wave-functional was interpreted as representing
the quantum black-hole behavior, mainly because the expected value
of the time derivative of the black-hole mass is negative. Furthermore,
the mass loss rate is in agreement with the one derived directly from
the Hawking emission process \cite{carr}.

In the next section \ref{sec:classical}, we re-write the Hamiltonian
form of the theory of general relativity for a spherically symmetric
space-time, minimally coupled to a massless scalar field. We use the
notation introduced by K. Kucha\v{r} in Ref. \cite{kuchar1}. We
obtain the total supermomentum and superhamiltonian for the 
gravitational
and matter sectors.

In Section \ref{sec:quantum}, we show that using Kucha\v{r}'s 
notation,
in the gauge proposed by Tomimatsu, the constraints are still 
proportional
to each other, although the proportionality constant is different
from Tomimatsu's. We present Tomimatsu's solutions to the 
Wheeler-DeWitt
equation and apply the causal interpretation for both 
wave-functionals.
We explicitly solve the dynamical equations for the mass variable
coming from both wave-functionals, to find the time dependence of
this variable. For the first wave-functional the mass increases with
time, representing the classical behavior, and for the second 
wave-functional
the mass decreases with time, representing the quantum behavior. The
time dependency for the mass loss rate is in agreement with predictions
on how the evaporation should take place, if one considers the 
\textit{elementary
particle} picture of black-hole emission \cite{carr}. We also compute
the quantum potential for both wave-functionals and confirm the classical
and quantum behavior of them.

Finally, in Section \ref{section:conclusion}, we summarize the main
points and results of the paper.

\section{Classical Formalism.}

\label{sec:classical}

As pointed out in the introduction we would like to study the quantum
general relativity theory of spherically symmetric, massless scalar
field minimally coupled to gravity. Therefore, our starting point
must be the Hamiltonian formalism for neutral, spherically symmetric
space-times, minimally coupled to a massless scalar field.

We start with the general, spherically symmetric metric, written in
the Arnowitt-Deser-Misner (ADM) form,

\begin{equation}
ds^{2}\,=\,-\, N^{2}dt^{2}\,+\,\Lambda^{2}\,(dr\,+\, N^{r}dt)^{2}\,+\, 
R^{2}d\Omega^{2}\,,\label{1}\end{equation}
 where $d\Omega^{2}$ is the metric on the unit two-sphere, and $N$,
$N^{r}$, $\Lambda$, and $R$ are functions of $t$ and $r$ only.
Here, we are using a unit system in which all physical constants are
set to the identity.

Next, we must write the action for the space-times given by the metric
(\ref{1}). The action $S$, for space-times with generic boundary
properties is given by the sum of an hypersurface term for the 
gravitational
sector ($S_{\Sigma}^{G}$), plus a boundary term for the gravitational
sector ($S_{\partial\Sigma}^{G}$), plus a term for the matter sector
\cite{wald1},

\begin{equation}
S\,=\,{\frac{1}{16\pi}}\,\int_{M}\, R\,(-g)^{1/2}\, dx^{4}\>+
\>{\frac{1}{8\pi}}\,\int_{\partial M}\, K\, h^{1/2}\, 
dx^{3}\,-\,{\frac{1}{8\pi}}\,\int_{M}\,
(-g)^{1/2}g^{\alpha\beta}\Phi,_{\alpha}\Phi,_{\beta}\, dx^{4}\,,
\label{2}
\end{equation}
 where $R$ is the curvature scalar, $g$ is the determinant of the
four-dimensional metric, $K$ is the trace of the second fundamental
form of the boundary, $h$ is the determinant of the three-dimensional
metric induced on the boundary, and $\Phi$ is the scalar field. Since
the boundary action $S_{\partial M}^{G}$ will not modify the equations
of motion, in the present case, we shall not consider it here. The
action of the gravitational sector will be entirely represented by
$S_{\Sigma}^{G}$.

Let us write down $S$ in terms of the fields describing the 
gravitational
degrees of freedom, $R$ and $\Lambda$, and their conjugate momenta.
In order to identify these momenta one must first cast the 
hypersurface
action in its ADM Lagrangian form \cite{wheeler}. Then, using the
results obtained in Ref. \cite{kuchar1}, we may easily write $S$
Eq. (\ref{2}), without $S_{\partial\Sigma}^{G}$, for the present
situation,

\begin{eqnarray}
S[R,\Lambda,\Phi;N,N^{r}]\>\! & = & \!\>\>\int\,
dt\,\int_{-\infty}^{\infty}\, dr\,\{\,{\frac{1}{N}}\,\{\,
R\,[\,(\Lambda N^{r})'\>-\>\dot{\Lambda}\,]\,(\,
N^{r}R'\>-\>\dot{R}\,)\nonumber \\ +\>{\frac{\Lambda}{2}}\,(\,
N^{r}R'\! & - & \!\dot{R}\,)^{2}\,\}\>+\>
N\,(\,{\frac{\Lambda}{2}}\>-\>{\frac{RR''}{\Lambda}}\>+
\>{\frac{RR'\Lambda'}{\Lambda^{2}}}\>-\>{\frac{R'^{2}}{2\Lambda}}\,)
\nonumber
\\ {\frac{1}{2}}\,[N^{-1}\Lambda R^{2}\! & ( & \!\dot{\Phi}\,-\,
N^{r}\Phi')^{2}\,-\, N\Lambda^{-1}R^{2}\Phi'^{2}]\,\}\,,\label{5}
\end{eqnarray}
where the over-dots and primes mean differentiations in the time and
radial parameters, respectively.

By functional differentiation of the above action Eq. (\ref{5}),
with respect to the velocities $\dot{\Lambda}$, $\dot{R}$ and 
$\dot{\Phi}$,
we obtain the momenta $P_{\Lambda}$, $P_{R}$ and $P_{\Phi}$,

\begin{eqnarray}
P_{\Lambda} & = & -\,{\frac{R}{N}}\,(\dot{R}\,-\, N^{r}R')\,,
\label{6}\\
P_{R} & = & -\,{\frac{1}{N}}\,\{\Lambda(\dot{R}\,-\, N^{r}R')\,+\, 
R[\dot{\Lambda}\,-\,(N^{r}\Lambda)']\}\,,\label{7}\\
P_{\Phi} & = & {\frac{\Lambda R^{2}}{N}}\,(\dot{\Phi}\,-\, 
N^{r}\Phi')\,.
\label{15}
\end{eqnarray}

Now, we are prepared to write the canonical Hamiltonian which has
the explicit form,

\begin{equation}
H_{c}\>=\> N\, H\>+\> N^{r}\, H_{r}\>,\label{9}\end{equation}
 such that \begin{eqnarray}
H\> & = & \>{\frac{\Lambda P_{\Lambda}^{2}}{2R^{2}}}\>-
\>{\frac{P_{R}P_{\Lambda}}{R}}\>+\>{\frac{RR''}{\Lambda}}\>-
\>{\frac{RR'\Lambda'}{\Lambda^{2}}}\>+
\>{\frac{R'^{2}}{2\Lambda}}\>-\>{\frac{\Lambda}{2}}\>+
\>{\frac{1}{2\Lambda}}\,(R^{-2}P_{\Phi}^{2}\,+\, 
R^{2}\Phi'^{2})\>,\label{10}\\
H_{r}\> & = & \> P_{R}R'\>-\>\Lambda P_{\Lambda}'\>+
\> P_{\Phi}\Phi'\>,
\label{11}
\end{eqnarray}
where $H_{r}$ and $H$ are, respectively, the supermomentum and
the superhamiltonian constraints of the model.

The hypersurface action is promptly written in terms of the canonical
Hamiltonian as

\begin{equation}
S[\Lambda,R,\Phi,P_{\Lambda},P_{R},P_{\Phi};N,N^{r}]\>=\>
\int dt\,\int_{-\infty}^{\infty}dr\,(P_{\Lambda}\dot{\Lambda}\,+\, 
P_{R}\dot{R}\,+\, P_{\Phi}\dot{\Phi}\,-\, NH\,-\, N^{r}H_{r})\>.
\label{12}
\end{equation}

\section{Quantum Formalism.}

\label{sec:quantum}

\subsection{Tomimatsu's Gauge.}

\label{subsec:tomimatsu}

In this section, we quantize the model described by the 
superhamiltonian
(\ref{10}). Since we want to study black-hole emission, we use the
gauge proposed by Tomimatsu in Ref. \cite{tomimatsu}. In this gauge,
we have,

\begin{equation}
N^{-2}\,=\,\Lambda^{2}\,=\,1\,+\,{\frac{2M}{r}}\,,\qquad 
N_{r}\,=\,{\frac{2M}{r}}\,,
\label{23}
\end{equation}
where $M=M(r,t)$ plays de role of a mass function. The apparent
horizon is located at,

\begin{equation}
r\,=\,2M(r,t)\,.
\label{24}
\end{equation}
If one imposes equation (\ref{24}) in the equation that determines
the apparent horizon,

\begin{equation}
g^{\alpha\beta}R,_{\alpha}R,_{\beta}\,,
\label{25}
\end{equation}
one finds that $R$ has to satisfy $R,_{t}=0$.

Near the apparent horizon, the gauge assumes that the massless scalar
filed $\Phi$ becomes ingoing and null. By introducing an advanced
time $v\equiv t+r$, one has the approximate forms

\begin{equation}
\Phi\,=\,\Phi(v)\,,\qquad M\,=\, M(v)\,,\qquad R\,=\, 
r\,,
\label{26}
\end{equation}
which will be valid near the apparent horizon.

Introducing all the above information coming from Tomimatsu's gauge
in the equations (\ref{6}), (\ref{7}), (\ref{15}) and (\ref{23}),
we obtain, respectively,

\begin{equation}
P_{\Lambda}\,=\,{\frac{R}{\sqrt{2}}}\,\qquad P_{R}\,=\,
{\frac{1}{2}}\dot{M}\,+\,{\frac{1}{4}}\,,\qquad P_{\Phi}\,=\, 
R^{2}\dot{\Phi}\,\qquad\Lambda\,=\,\sqrt{2}\,,
\label{27}
\end{equation}
where over-dot means derivative with respect to $v$.

We notice that, from equation (\ref{27}), $\Lambda$ cannot be 
considered
a dynamical variable. It means that the variables describing 
the model
after the imposition of Tomimatsu's gauge will be $R$ and $\Phi$.
$R$ is a variable because, from equations (\ref{24}) and (\ref{26}),
it determines the position of the apparent horizon through the 
relationship

\begin{equation}
R\,=\,2M(v)\,.
\label{28}
\end{equation}

Now, we may compute the value of $H$ and $H_{r}$, from equations
(\ref{10}) and (\ref{11}), respectively, in Tomimatsu's gauge. With
the aid of equation (\ref{27}), we obtain that,

\begin{equation}
H\,=\,{\frac{H_{r}}{\sqrt{2}}}\,=\,\sqrt{2}\,
\left(\,{\frac{P_{\Phi}^{2}}{2R^{2}}}\,-\, P_{R}\,+\,
{\frac{1}{4}}\,\right)\,.
\label{29}
\end{equation}
The constraints are still proportional to each other, although the
proportionality constant is different from Tomimatsu's \cite{tomimatsu}.

\subsection{Wave-function.}

\label{subsection:wavefunction}

We would like to quantize the theory using Dirac's formalism 
for quantizing
constrained systems \cite{dirac}. First, we introduce a 
wave-function
which is a functional of the canonical fields in their operatorial
form $\hat{R}$ and $\hat{\Phi}$,

\begin{equation}
\Psi\>=\>\Psi[\,\hat{R},\,\hat{\Phi}\,]\>.\label{30}\end{equation}
Then, we impose the appropriated commutators between the fields 
operators
and their conjugated momenta $\hat{P}_{R}$, $\hat{P}_{\Phi}$. 
Finally,
we demand that the operatorial form of the constraints, equation 
(\ref{29}),
annihilate the wave-function equation (\ref{30}).

Working in the fields representation the operators $\hat{R}$ and
$\hat{\Phi}$ are replaced by the fields themselves, and the conjugate
momenta are defined as the following functional derivatives,

\begin{equation}
\hat{P}_{R}\>=\>-\, i\,{\frac{\delta}{\delta R}}\>,
\label{31}
\end{equation}
\begin{equation}
\hat{P}_{\Phi}\>=\>-\, i\,{\frac{\delta}{\delta\Phi}}\>,
\label{32}
\end{equation}
where we are using units where $\hbar=1$.

The most important motivation for Tomimatsu's gauge is the result
that $H$ is proportional to $H_{r}$. It means that one has to consider
only one of the constraints. We may write the operatorial expression
of the constraint equation (\ref{29}) in terms of the new variable
$T\equiv1/R$, and the operatorial expressions for the momenta 
$\hat{P}_{R}$
(\ref{31}), and $\hat{P}_{\Phi}$ (\ref{32}). If we demand that
$\Psi$ in equation (\ref{30}) satisfies the operatorial constraint
equation, we obtain, 

\begin{equation}
-\,{\frac{1}{2}}{\frac{\partial^{2}\Psi}{\partial\Phi^{2}}}\,-\, 
i{\frac{\partial\Psi}{\partial T}}\,+\,{\frac{1}{4T^{2}}}\Psi\,=\,
0\,.
\label{33}
\end{equation}

Now, from Ref. \cite{tomimatsu} we have the following solutions to
equation (\ref{33}),

\begin{equation}
\Psi_{c}\,=\, C\exp{\left[i\left({\frac{1}{4T}}\,-\,
{\frac{1}{2}}k^{2}T\,+\, k\Phi\right)\right]}\,,
\label{34}
\end{equation}
and 
\begin{equation}
\Psi_{q}\,=\, C\exp{\left[i\left({\frac{1}{4T}}\,+\,
{\frac{1}{2}}k^{2}T\right)\,-\,|k\Phi|\right]}\,,
\label{35}
\end{equation}
where $k$ and $C$ are arbitrary real and complex parameters, 
respectively.

Tomimatsu concluded that $\Psi_{c}$ equation (\ref{34}) represents
the classical black-hole behavior \cite{tomimatsu}. If one computes
the expectation value of $\dot{M}$: $\langle\dot{M}\rangle\,=\,
\langle2P_{R}-1/2\rangle$,
one finds a positive value. It means that the apparent horizon 
increases
and the black-hole can only absorb. Also the scalar field sector is
described by scalar waves penetrating the apparent horizon from the
exterior region.

On the other hand, $\Psi_{q}$ in equation (\ref{35}) represents
the quantum-mechanical black-hole behavior \cite{tomimatsu}. In this
case the value of $\langle\dot{M}\rangle$ is given by

\begin{equation}
\langle\dot{M}\rangle\,=\,-\,{\frac{k^{2}}{4M^{2}}}\,.
\label{36}
\end{equation}
The rhs of equation (\ref{36}) is always negative, which means that
the apparent horizon decreases and the black-hole can only emit. The
scalar field cannot penetrate the horizon; it is exponentially 
suppressed.
This can be interpreted as a classically forbidden state.

\subsection{Causal Interpretation.}

\label{subsection:causal}

Let us see, in the present subsection, what the causal 
interpretation
tell us about the states described by the wave-functionals in 
equations
(\ref{34}) and (\ref{35}). Following the causal interpretation 
formalism
applied to quantum general relativity \cite{acacio}, if we write
our wave-functionals in equations (\ref{34}) and (\ref{35}) as,

\begin{equation}
\Psi\,=\,{\mathcal{R}}\,\exp{(iS)}\,,
\label{37}
\end{equation}
we may obtain a dynamical equation for the physical variables in
the following way,

\begin{equation}
P_{X_{i}}\,=\,{\frac{\delta S}{\delta X_{i}}}\,,
\label{38}
\end{equation}
where $X_{i}$ stands for $R$ and $\Phi$. Also, there is a quantum
potential $Q$, which governs the dynamics of the system. 
The expression for $Q$ is given, in the present situation, by,

\begin{equation}
Q\,=\,-\,{\frac{\nabla^{2}{\mathcal{R}}}{{\mathcal{R}}}}\,.
\label{39}
\end{equation}
The dynamical equations for $\Phi$ for both wave-functionals 
(\ref{34})
and (\ref{35}) are trivial and do not bring any contribution 
to the
understanding of the system. Therefore, we shall restrict our 
attention to the dynamical equation for $R$.

Starting with $\Psi_{c}$ equation (\ref{34}), we may write the 
dynamical equation (\ref{38}) for $X_{i}=R$,

\begin{equation}
P_{R}\,=\,{\frac{1}{4}}\,+\,{\frac{k^{2}}{2R^{2}}}\,.
\label{40}
\end{equation}
Now, introducing the expression of $P_{R}$ given in 
equation (\ref{27})
in equation (\ref{40}), we obtain the following equation for the
evolution of $M$,

\begin{equation}
\dot{M}\,=\,{\frac{k^{2}}{4M^{2}}}\,.
\label{41}
\end{equation}
This equation is easily integrated to give,

\begin{equation}
M^{3}\,=\,{\frac{3}{4}}k^{2}(v\,-\, v_{0})\,+\, M_{0}^{3}\,,
\label{42}
\end{equation}
where $v_{0}$ and $M_{0}$ are the initial values of $v$ and $M$,
respectively.

Solution (\ref{42}) tell us that the black-hole mass M increases
continuously as the time, measured by $v$, increases. This 
wave-functional
is associated with the classical behavior of the black-hole. 
In particular,
if we compute the value of the quantum potential $Q$ from equation
(\ref{39}) for $\Psi_{c}$ in equation (\ref{34}), we find that
it is zero, as expected for the classical situation.

Consider, now, the dynamical equation for $M$ coming from $\Psi_{q}$
equation (\ref{35}). With the aid of $P_{R}$ in equation (\ref{38}),
which in this case is

\begin{equation}
P_{R}\,=\,{\frac{1}{4}}\,-\,{\frac{k^{2}}{2R^{2}}}\,,
\label{43}
\end{equation}
and $P_{R}$ from equation (\ref{27}), we find the dynamical equation
for $M$,

\begin{equation}
\dot{M}\,=\,-\,{\frac{k^{2}}{4M^{2}}}\,.
\label{44}
\end{equation}
Note that equation (\ref{44}) is similar to equation (\ref{36})
for the expectation value of $\dot{M}$. The difference is that 
equation
(\ref{44}) can be integrated to give the exact evolution of $M$
and not just the expectation value of this evolution. This 
equation is easily integrated to give

\begin{equation}
M^{3}\,=\,-\,{\frac{3}{4}}k^{2}(v\,-\, v_{0})\,+\, M_{0}^{3}\,,
\label{45}
\end{equation}
where $v_{0}$ and $M_{0}$ are the initial values of $v$ and $M$,
respectively. Equation (\ref{45}) tell us that if the black-hole
has an initial mass $M_{0}$ at $v_{0}$ after a time 
$v_{e}=4M_{0}^{3}/3k^{2}+v_{0}$,
it will completely evaporate. This is in accordance with the 
qualitative
predictions made by S. W. Hawking that, taking into account quantum
properties, black-holes evaporate \cite{hawking}. Equation (\ref{45})
is also in accordance with predictions on how this evaporation should
take place, if one considers the \textit{elementary particle} picture
of black-hole emission \cite{carr}. The quantum potential $Q$ in
equation (\ref{39}), computed for $\Psi_{q}$ in equation (\ref{35}),
is given by $-k^{2}/2R^{2}$. This may be interpreted as an attractive
potential that pulls $R$ to zero. If we remember that $R=2M$, this
potential also pulls the mass towards zero.

\section{Conclusions.}

\label{section:conclusion}

In this work we applied the de Broglie-Bohm interpretation of quantum
mechanics, also known as the causal interpretation, to the quantized
spherically symmetric black-hole coupled to a massless scalar field.
The wave-functional used was first obtained by Tomimatsu using the
standard ADM quantization and a gauge that places the observer close
to the black-hole horizon. In Tomimatsu's paper, he obtains two 
wave-functionals
that are solutions to the Wheeler-DeWitt equation, one of which predicts
a decreasing expected value for the black-hole mass and another that
predicts the standard classical result. Using the causal interpretation,
we computed the individual quantum trajectories determined by the
initial conditions. We showed that the quantum trajectories for the
black-hole mass could either increase or decrease with time. The quantum
trajectories that show increasing mass represent the usual black-hole
behavior of continuous energy absorption. The mass-decreasing quantum 
trajectories are a purely quantum mechanical phenomena. They can 
be physically interpreted as describing a black-hole that evaporates.

\acknowledgements

J. Acacio de Barros and G. Oliveira-Neto would like to thank FAPEMIG
and T. B. Vale would like to thank CAPES for the financial support.

\end{document}